\documentclass{fic-l}

\newtheorem{theorem}{Theorem}[section]

\theoremstyle{definition}

\theoremstyle{remark}

\numberwithin{equation}{section}

\begin{document}

\title{Perturbative Algebraic Field Theory, 
and Deformation Quantization}

\author{Michael D\"utsch}

\thanks{Work supported by the Deutsche 
Forschungsgemeinschaft.} 
\author{Klaus Fredenhagen}
\address{
II. Institut f\"ur Theoretische Physik\\
Universit\"at Hamburg\\
Luruper Chaussee 149\\
D-22761 Hamburg, Germany.}

\email{\tt duetsch@mail.desy.de, fredenha@x4u2.desy.de}

\date{January 11, 2001}

\dedicatory{Dedicated to Sergio Doplicher and John Roberts 
at the occasion of their 60th anniversary}

\dedicatory{Dedicated to Sergio Doplicher and John Roberts 
at the occasion of their 60th anniversary}

\begin{abstract}
A perturbative formulation of algebraic field theory is 
presented, both for the classical and for the quantum case, 
and it is shown that the relation between them may be 
understood in terms of deformation quantization. 
\end{abstract}

\maketitle

\section{Introduction}
The algebraic approach to field theory (``Local Quantum 
Physics'') \cite{B:haag} has 
deepened and enlarged our understanding of fundamental 
properties of quantum field theory \cite{B:bu}. As the perhaps most 
important insight one may mention the theory of 
superselection sectors \cite{B:dhr} which culminated in the work of 
Sergio Doplicher and John Roberts on a new duality 
theory for compact groups \cite{B:dr1,B:dr2}. 

On the level of concrete models the algebraic approach was 
less successfull. The unfortunately still unsufficient 
mathematical control on models of quantum field theory in all 
existing approaches seemed to be a severe obstacle for an 
application of the framework of algebraic field theory. 
But recently it was shown that on the 
level of perturbation theory a quite satisfactory 
formulation is possible, on the basis of older attempts 
by Bogoliubov-Shirkov \cite{B:bos}, Epstein-Glaser \cite{B:eg}, 
Steinmann \cite{B:steinmann} and Stora \cite{B:stora}. The main 
new insight is that the formulation of perturbative quantum 
field theory in the spirit of local 
quantum physics admits a complete disentanglement of 
ultraviolet and infrared problems. One application is the 
perturbative renormalization on a curved background 
\cite{B:bfroma,B:bf}, another a local construction of 
observables in gauge theories \cite{B:df1}.  A third 
application is a better understanding of the relation 
between classical and quantum field theory. It can be shown 
that the formalism of deformation quantization can be 
naturally applied and delivers the loop expansion of 
perturbation theory \cite{B:df2}. 

The plan of this paper is as follows: we briefly review the 
framework of algebraic quantum field theory and describe 
an analogous formulation of classical field theory. In 
particular we present a perturbative construction of 
interacting classical fields. 

We then discuss the problem of $*$-product quantization for 
free fields. We 
will show that the Wick quantization plays a distinguished 
role by allowing an extension to polynomials of fields. 

Up to singularities at coinciding points, the expansion of 
interacting classical fields in terms of iterated retarded 
Poisson brackets can be transformed into the expansion of the 
interacting quantum fields in retarded functions just by 
replacing Poisson brackets by commutators. The fixing of the 
ambiguities at coinciding points amounts to imposing 
renormalization conditions.    
\section{The Algebraic Formulation of Quantum Field Theory}

Algebraic quantum field theory essentially relies on 2 
principles:
\begin{enumerate}
        \item  Quantum principle: The observables form a 
        noncommutative associative $*$-algebra with a faithful 
        Hilbert space representation (e.g. a C*-algebra). 

        \item  Principle of locality: Observables are associated to 
        space time regions.
\end{enumerate}
The second principle allows an interpretation of 
measurements. If ${\mathfrak{A}}({\mathcal{O}})$ denotes the 
algebra of all observables which can be measured within the 
spacetime region ${\mathcal{O}}$ then the guiding principle 
of algebraic quantum field theory states that the isotonic,
 \begin{displaymath}
        {{\mathcal{O}}}_{1} \subset {{\mathcal{O}}}_{2}\Rightarrow 
        {\mathfrak{A}}({{\mathcal{O}}}_{1})  \hookrightarrow
        {\mathfrak{A}}({{\mathcal{O}}}_{2})\ ,
 \end{displaymath}
net $({\mathfrak{A}}({\mathcal{O}}))_{{\mathcal{O}}}$ 
characterizes the theory completely. This principle, 
originally formulated by Haag and Kastler in 
\cite{B:hk}, has been checked in a huge variety of 
situations. In particular it emphasizes that the physical 
interpretation of the theory does not depend on the choice 
of fields. This was already known at that time for 
scattering theory (fields from the same Borchers class 
produce the same $S$-matrix \cite{B:borchers1}). 
It plays a fundamental role in 
the analysis of superselection sectors 
\cite{B:borchers2,B:dhr,B:bufred,B:frs,B:dr2}, and is crucial for 
the approach of Buchholz and Verch towards an intrinsic 
renormalization group \cite{B:bv}. 

If one adopts this principle one gets a convenient notion 
of equivalence between different theories: two theories are 
equivalent if and only if their local nets are isomorphic. 
This notion of equivalence can be applied to an analysis of duality 
transformations and was recently used in Rehren's work 
\cite{B:rehren} on 
the relation between theories on D+1-dimensional Anti-de-Sitter spacetime 
and conformal field theories on D-dimensional Minkowski 
space originally conjectured by Maldacena in the framework 
of string theories.       
\section{The Algebraic Formulation of Classical Field Theory}
Let ${\mathcal{L}}(\varphi,\partial\varphi)$ be the Lagrangian of a scalar 
field $\varphi$ leading to the field equation
\begin{equation}
        \partial_{\mu}\frac{\partial{\mathcal{L}}}{\partial(\partial_{\mu}\varphi)} 
        =\frac{\partial{\mathcal{L}}}{\partial\varphi}\ ,\label{fieldeq}
\end{equation}
and let $\mathcal{C}$ be the space of smooth solutions with 
compactly supported Cauchy data. $\mathcal{C}$ may be 
considered as the classical phase space, and typical 
observables are the evaluation functionals 
$F_{x}(\varphi):=\varphi(x)$. By the usual abuse of notation we
write $\varphi(x)$ for $F_{x}$.
Following Peierls \cite{B:peierls}, one may define the 
Poisson bracket of two observables without recourse to a 
Hamiltonian formulation in the following way. Let 
${{\mathcal{L}}}_{1}$ be a polynomial in $\varphi$ (i.e. in $F_{x}$)
and $f$ be a 
test function on Minkowski space ${\mathbb{M}}$ with compact support. 
Denote by $\varphi_{f{{\mathcal{L}}}_{1}}$ a 
solution of the field equation (\ref{fieldeq})
derived from the Lagrangian ${{\mathcal{L}}}+f{{\mathcal{L}}}_{1}$. 
Then $\varphi_{f{{\mathcal{L}}}_{1}}$ coincides at early times with 
a solution $\varphi_{\mathrm{in}}$ of the original field 
equation and at late times with another solution 
$\varphi_{\mathrm{out}}$. Provided the Cauchy problem is well 
posed, we obtain a mapping
\begin{displaymath}
        s(f{{\mathcal{L}}}_{1}):\left\{
        \begin{array}{ccc}
                \mathcal{C} & \to & \mathcal{C}  \\
                \varphi_{\mathrm{in}} & \mapsto & \varphi_{\mathrm{out}}
        \end{array}\right. 
\end{displaymath}
which may be considered as the classical $S$-matrix. The 
Poisson bracket between the observables 
${{\mathcal{L}}}_{1}(F_{x})$ with an arbitrary observable $G$ 
is now defined by
\begin{displaymath}
        \{{{\mathcal{L}}}_{1}(F_{x}),G\}(\varphi)
        \overset{\mathrm{def}}{=}\frac{\delta}{\delta f(x)} 
        G\bigl( s(f{{\mathcal{L}}}_{1})^{-1}\varphi\bigr)\vert_{f=0}\ .
\end{displaymath}
If we, for example, take the free Klein-Gordon field and choose 
${{\mathcal{L}}}_{1}=\varphi$, then the interacting field 
\begin{equation}
        \varphi_{f{{\mathcal{L}}}_{1}}=\varphi - 
        \Delta_{\text{ret}}*f \label{phi}
\end{equation}
solves (\ref{fieldeq}),
where $\Delta_{\text{ret}}$ is the retarded solution of 
the Klein-Gordon equation and $*$ denotes convolution.
The incoming field coincides with the original free field $\varphi$, 
and the outgoing field is\footnote{Obviously the expression (\ref{phi-out})
fulfills the free field equation and agrees for late times with 
$\varphi_{f{{\mathcal{L}}}_{1}}$ (\ref{phi}).}
\begin{equation}
        \varphi_{\mathrm{out}}=\varphi -\Delta *f\ ,\label{phi-out}
\end{equation}
where $\Delta = \Delta_{\text{ret}}-\Delta_{\text{adv}}$ is 
the commutator function from quantum field theory. For the 
Poisson brackets of the classical fields $\varphi(x)$ one finds
\begin{displaymath}
        \{\varphi(x),\varphi(y)\}=\Delta(x-y)\ .
\end{displaymath}
One may now start from the Poisson algebra of the free 
field $\varphi$ and construct interacting fields perturbatively. So 
let ${\mathcal{L}}_{1}$ be an arbitrary polynomial and $f$ a test 
function. The corresponding interacting field is given by 
the following formal power series in the coupling constant 
$f\in {\mathcal D}(\mathbb{M})$
\begin{gather}
        \varphi_{f{\mathcal{L}}_{1}}(x) = \sum_{n\ge 0}\int_{x^{0}\ge 
        x_{1}^{0}\ge \ldots \ge x_{n}^0}{\mathrm{d}} x_{1}\ldots 
        {\mathrm{d}} x_{n} \notag\\
        \cdot f(x_{1})\cdots f(x_{n}) 
        \{{\mathcal{L}}_{1}(x_{n}),\{\ldots\{{\mathcal{L}}_{1}(x_{1}),\varphi(x)\}\ldots\}\}
\label{intfield:class}
\end{gather}
The integrals are well defined because of the support 
restrictions on $f$. One verifies that 
$\varphi_{f{\mathcal{L}}_{1}}(x)$ solves the 
field equation (\ref{fieldeq}), and that for equal times
$\varphi_{f{\mathcal{L}}_{1}}$ and 
$\dot{\varphi}_{f{\mathcal{L}}_{1}}$ 
have canonical Poisson brackets \cite{B:df2,B:df3}. 
Moreover, when $f\equiv 1$ within 
the causally complete region ${\mathcal{O}}$, the Poisson 
algebra generated by the interacting field  $\varphi_{f{\mathcal{L}}_{1}}(x) $ 
with $x\in {\mathcal{O}}$ is, up to canonical 
transformations, independent of $f$ (see section 5). Hence we directly 
obtain an inductive system of local Poisson algebras of 
classical observables, which we may consider as the 
classical theory.    
\section{Wick quantization}
In deformation quantization \cite{B:bffls} one studies a family 
$*_{\hbar}$ of associative products in a given Poisson 
algebra such that
\begin{displaymath}
        \lim_{\hbar\to 0}a*_{\hbar}b=ab\ ,\ \lim_{\hbar\to 0} 
        \frac{1}{i\hbar}[a,b]_{\hbar}=\{a,b\}\ .
\end{displaymath}
In the case of free field theory one may choose
\begin{displaymath}
        \varphi(x)*_{\hbar}\varphi(y)= 
        \varphi(x)\varphi(y)+\hbar\Delta'(x-y)
\end{displaymath}
where the antisymmetric part  of $\Delta'$ is determined by 
the commutator function,
\begin{equation}
         \Delta'(x)-\Delta'(-x)=i\Delta(x)\ .\label{antisym-part}
\end{equation}
For (classical) products of fields one may set
\begin{displaymath}
   \varphi(x_{1})\cdots \varphi(x_{n}) *_{\hbar} \varphi(y_{1})\cdots 
   \varphi(y_{m}) = 
\end{displaymath}  
\begin{equation}
   \sum_{\text{contractions}}\prod \hbar\Delta'(x_{i}-y_{k}) \prod 
   \varphi(x_{j})\varphi(y_{l})\ ,\label{Wick}
\end{equation}
with the combinatorics known from Wick's theorem (see (\ref{Wick:erzF})).
There are several possibilities for the choice of $\Delta'$. 
The most symmetric one is $\Delta'=\frac{i}{2}\Delta$. This 
choice 
leads to the Weyl-Moyal quantization. Another choice is 
$\Delta'=\Delta^+$, where $\Delta^+$ is the positive 
frequency part of $i\Delta$. Then one gets the so called Wick 
quantization. 

On the level of the algebra generated by the smeared 
fields $\varphi(f)=\int{\mathrm{d}} x \varphi(x)f(x)$ the different choices 
lead to isomorphic algebras. Namely, there is a 
differentiable family 
$T_{\hbar}$ of linear invertible maps, interpolating between 
the different $*$-products, with $T_{0}=1$. To see this we 
formulate our $*$-products in terms of generating 
functionals. We interprete $e^{\varphi(f)}$ as the generating functional 
for the products of classical fields. Therewith, the formula (\ref{Wick})
can be written in the form
\begin{equation}
    e^{\varphi(f)}*_{\hbar} e^{\varphi(g)}=e^{\varphi(f+g)}\label{Wick:erzF}
    e^{\hbar(f,\Delta' g)}
\end{equation}
where $(f,\Delta' g)=\int{\mathrm{d}} x{\mathrm{d}} y 
f(x)\Delta'(x-y)g(y)$. 
We set
\begin{displaymath}
        T_{\hbar}(e^{\varphi(f)})\overset{\mathrm{def}}{=} e^{\varphi(f)} 
        e^{\frac{\hbar}{2}(f,(\Delta''-\Delta')f)}\ .
\end{displaymath}
Then $T_{\hbar}$ interpolates between $*$-products defined 
with $\Delta''$ and $\Delta'$, respectively:
\begin{displaymath}
T_{\hbar}(e^{\varphi(f)})*_{\hbar}^{\Delta''}T_{\hbar}(e^{\varphi(g)})=
T_{\hbar}(e^{\varphi(f)}*_{\hbar}^{\Delta'}e^{\varphi(g)}),
\end{displaymath}
where we use $(f,(\Delta''-\Delta') g)=(g,(\Delta''-\Delta')f)$
(which relies on (\ref{antisym-part})).

But if we go beyond this minimal algebra and include also 
pointwise products of fields (this is necessary for a 
description of interesting interactions) then the picture 
changes. We now can accept only functions $\Delta'$ for 
which the products can be defined at coinciding points.
We may look, for example,  at the products 
of $\varphi(x)^2$. We obtain
\begin{displaymath}
        \varphi(x)^2 *_{\hbar} \varphi(y)^2 = \varphi(x)^2\varphi(y)^2 +2 
        \Delta'(x-y)\varphi(x)\varphi(y) + 2 \Delta'(x-y)^2
\end{displaymath}
which makes sense only when the square of $\Delta'$ can be 
defined.

A convenient criterion for the existence of products of 
distributions or, equivalently, for the existence of 
restrictions of tensor products of distributions to 
submanifolds of coinciding points can be formulated in terms of the 
wave front sets \cite{B:micro}. Namely, a distribution can be restricted to 
a submanifold, if the conormal bundle of the submanifold 
does not intersect the wave front set of the distribution. 
The wave front set of the commutator function is
\begin{displaymath}
        {\mathrm{WF}}(\Delta)=\{(x,k)\in T^{*}{\mathbb{M}}, k\ne 0, x,k 
        \text{ lightlike }, k \text{ coparallel to } x\}
\end{displaymath}
that of $\Delta_{+}$ is the positive frequency part 
$k_{0}>0$ of 
the wave front set of $\Delta$ (see e.g. \cite{B:radman}). 

In our example we have to study the conormal bundle $N^{*}D$ of 
the diagonal $D=\{(x,x),x\in{\mathbb{M}}\}$ of ${\mathbb{M}}^2$. It is the 
orthogonal complement of the tangent bundle of $D$ within 
$T^{*}_{D}{\mathbb{M}}^2$, 
\begin{displaymath}
        N^{*}D=\{(x,x;k,-k), (x,k)\in T^{*}{\mathbb{M}}\}\ .
\end{displaymath}
It certainly cannot intersect the wave front set of 
$\Delta_{+}\otimes \Delta_+$ because of the positive 
frequency condition whereas there is a nontrivial 
intersection with the wave front set of $\Delta\otimes 
\Delta$. 

It is now straightforward to see that the $*$-product with 
$\Delta'=\Delta_{+}$ can be extended to the Poisson algebra 
containing all smeared powers $\varphi^n(f)=\int{\mathrm{d}} x 
\varphi^n(x)f(x)$ of the free field $\varphi$. The mappings $T$ to 
equivalent $*$-products are 
only well defined on this larger Poisson algebra if 
$\Delta'$ differs from $\Delta_{+}$ by 
a smooth function. The $*$-product with 
$\Delta'=-\Delta_{-}$, $\Delta_{-}$ being the negative 
frequency part of $i\Delta$, could also be defined on the 
larger Poisson algebra. But on the arising associative 
algebra there is no linear 
functional $\omega$ with $\omega(\mathbf{1})=1$ 
which is nonnegative on $\varphi(f)*_{\hbar}\varphi(f)$ for all real 
valued test functions $f$. Namely, we have
\begin{displaymath}
        \omega(\varphi(f)*_{\hbar}\varphi(f)) = 
        \omega(\varphi(f)^2) +\hbar (f,\Delta' f)\ .
\end{displaymath}
If $f$ tends to the $\delta$-function, the first term on 
the right hand side 
converges whereas the second term tends to $\pm\infty$ for 
$\Delta'=\pm \Delta_{\pm}$. 
We conclude that the algebra in the case of $\Delta'=-\Delta_{-}$ 
does not admit a faithful 
Hilbert space representation with a hermitean field $\varphi$.

We observe that, in contrast to quantum mechanics with 
finitely many degrees of freedom, Wick quantization is distinguished in 
field theory (see also \cite{B:di}).

We may now formalize the structure described above. The 
admissable smearing functions of $n$-fold products of the 
free field are symmetrical distributions $t_{n}$ with compact 
support and with a wave front set where never all components 
of the covectors $k$ are contained in the closure of the same 
component of the lightcone (forward 
or backward)
\begin{displaymath}
        k\not\in \overline{V}_{+}^n \cup \overline{V}_{-}^n\ .
\end{displaymath}
The space of all these distributions will be denoted by 
${\mathcal{W}}_{n}$. It contains in particular products of a 
$\delta$-function in the difference variables with a smooth 
function of compact support (cf. \cite{B:df2}).

The $*$-product may directly be defined in terms of these 
smearing functions. Let ${\mathcal{W}}_0 ={\mathbb{C}}$ and
${\mathcal{W}}=\bigoplus_{n}{\mathcal{W}}_{n}$, 
and let $f_{n}$ denote the component of 
$f\in {\mathcal{W}}$ in ${\mathcal{W}}_{n}$. 
Then we define an associative 
product $*_{\hbar}$ on ${\mathcal{W}}$ by
\begin{displaymath}
     (t*_{\hbar}s)_{n}=\sum_{n+2k=l+m}\hbar^k 
        t_{m}\otimes_{k}s_{l}\ .
\end{displaymath}
Here $\otimes_{k}$ denotes the $k$-times, with $\Delta_{+}$, contracted tensor 
product. This is the symmetrical distribution, which is 
defined on symmetrical test functions $f\in {\mathcal D}(\mathbb{M}^{m+l-2k})$
$(m\geq k, l\geq k)$ by
\begin{displaymath}
        \langle t_{m}\otimes_{k}s_{l} , f\rangle 
        = \frac{m!l!}{k!(m-k)!(l-k)!} 
        \langle 
        t_{m}\otimes s_{l}, (\hat{\Delta}_{+}^{\otimes k} \otimes 
        f)\circ \sigma \rangle
\end{displaymath}
where $\hat{\Delta}_{+}(x,y)=\Delta_{+}(x-y)$ and where 
$\sigma$ permutes the components of the coordinates of 
$(x,y)\in {\mathbb{M}}^m\times {\mathbb{M}}^l$ such that
\begin{displaymath}
        \sigma(x_{1},\ldots,x_{m},y_{1},\ldots,y_{l})= 
        (x_{1},y_{1},\ldots, 
        x_{k},y_{k},x_{k+1},\ldots,x_{m},y_{k+1},\ldots ,y_{l})\ .
\end{displaymath}
It is easy to see that the product is well defined,  
satisfies the condition on the wave front set and makes 
${\mathcal{W}}$ to an associative algebra.  
The relation of this abstractly defined algebra with the 
algebra of smeared Wick products on Fock space is
described in the following Theorem (cf. \cite{B:df2}):
 \begin{theorem}
        Let $\phi$ be the mapping from ${\mathcal{W}}$ into the 
        the space of densely defined operators on Fock space 
        \begin{displaymath}
                \phi(t)=\sum_{n}:\!\varphi^{\otimes n}\!:(t_{n})
        \end{displaymath}
        with the Wick products $:\!\varphi^{\otimes 
        n}\!:(x_{1},\ldots,x_{n})= :\!\varphi(x_{1})\cdots 
        \varphi(x_{n})\!:,\>\varphi^{\otimes 0}= \mathbf{1}$. 
        Then $\phi$ is an algebra homomorphism with the kernel
        \begin{displaymath}
                \operatorname{Ke}{\phi}= 
        \{t,\exists s\in{\mathcal{W}} \text{ such that } 
                t_{n}=\sum_{i}(\partial^\nu_{i}\partial_{i\,\nu} 
                +m^2)s_{n}\quad \forall n\}.
        \end{displaymath}
 \end{theorem} 
So, $\bigl( (\mathcal{W}/{\operatorname{Ke}{\phi}}),*_\hbar\bigr)$ 
provides a purely 
algebraic quantization of the given Poisson algebra of the classical free 
fields, it expresses the algebraic structure of smeared Wick products 
without using the Fock space.
Starting from ${\mathcal{W}}$, the Fock representation is induced by the state 
 \begin{displaymath}
        \omega_{0}:\left\{
        \begin{array}{ccc}
                {\mathcal{W}} & \to & {\mathbb{C}}  \\
                t & \mapsto & t_{0}
        \end{array}\right. \ .
 \end{displaymath}
via the GNS-construction, in particular it holds
 \begin{displaymath}
        \omega_{0}(t) =(\Omega,\phi(t)\Omega)
 \end{displaymath}
 with $\Omega$ the vacuum vector in Fock space (see also \cite{B:bw}). 
 \section{Loop expansion and deformation quantization}
 Formally, we obtain the interacting quantum field by 
 replacing in the formula for the interacting classical field
 (\ref{intfield:class}) the 
 Poisson bracket by the commutator with respect to the 
 associative product $*_{\hbar}$, divided by $i\hbar$ (cf. \cite{B:df2}).
 So we do not deform directly the algebra of the perturbative interacting 
 classical fields, instead we deform the 
 underlying Poisson algebra of free fields.
 For a polynomial interaction, the quantum field becomes a 
 convergent\footnote{Convergence is meant here in the sense of formal
power series in the coupling constant 
$g\in {\mathcal D}({\mathbb{M}})$, in formula:
$\forall N\in{\mathbb{N}}$ there exists $M\in{\mathbb{N}}$ 
such that $\sum_{k=n}^m \hbar^k
 \varphi_{g{\mathcal{L}}_{1}}^{(k)}={\mathcal O}(g^N)\quad\forall n,m\geq M$.}
 power series in $\hbar$,
 \begin{displaymath}
\varphi_{g{\mathcal{L}}_{1}}^{\hbar} = 
\sum \hbar^n \varphi_{g{\mathcal{L}}_{1}}^{(n)}\ .
 \end{displaymath}
 The $n$-th term is just the $n$-loop contribution in an 
 expansion into Feynman diagrams. 
 
 Let us introduce the algebra of functions of $\hbar$ with 
 values in the formal power series over $g\in{\mathcal{D}}({\mathbb{M}})$ 
 with coefficients in ${\mathcal{W}}$,
 \begin{displaymath}
        \mathcal{V}= \{A:\mathbb{R}_{+} \to \mathcal{W}[[g]]\}\ ,
 \end{displaymath}
 with the product
 \begin{displaymath}
      (A*B)(\hbar) \overset{\mathrm{def}}{=} A(\hbar) *_{\hbar} B(\hbar) \ .
 \end{displaymath}
 The interacting fields generate a subalgebra ${\mathfrak{A}}$ of 
 ${\mathcal{V}}$.
 ${\mathfrak{A}}$ may be considered as an algebra of sections of a 
 bundle of algebras $\mathfrak{A}_{\hbar}$ over $\mathbb{R}_{+}$. The 
 elements of order $\hbar$ generate an ideal 
${\mathfrak{I}}$ whithin ${\mathfrak{A}}$ 
 which is also an ideal with respect to the Poisson bracket
 \begin{equation}
      \{A,B\}(\hbar)\overset{\mathrm{def}}{=} 
      \frac{1}{i\hbar}[A,B](\hbar)\ .\label{Pb}
 \end{equation}
 The quotient ${\mathfrak{A}}_{0}={\mathfrak{A}} /{\mathfrak{I}}$ is the 
classical Poisson 
 algebra. The powers ${\mathfrak{I}} ^n,\>n\in{\mathbb{N}},$ 
generate the ideals
whithin ${\mathfrak{A}}$ of the elements of order $\hbar^n$, and
the algebras of observables up to $n$ loops can be defined by
 \begin{equation}
        {\mathfrak{A}}_{n}\overset{\mathrm{def}}{=} 
        {\mathfrak{A}} / {\mathfrak{I}}^{n+1}\label{A_n}
 \end{equation}
(cf. \cite{B:df2}).
 They are noncommutative algebras with an additional Poisson 
 bracket (\ref{Pb}) which is not defined in terms of the commutator (it 
 comes from the commutator at higher order in $\hbar$). The 
projective system
\begin{displaymath}
        \mathfrak{A}\rightarrow ...\rightarrow \mathfrak{A}_{n+1}
\rightarrow \mathfrak{A}_n\rightarrow...\rightarrow \mathfrak{A}_0
 \end{displaymath}
interpolates between the quantum theory $\mathfrak{A}$ and the classical
theory $\mathfrak{A}_0$.

 To make the described reasoning rigorous, one has to treat 
 infrared and ultraviolet problems. The infrared problems 
 are in our approach circumvented, in the first step, by 
 allowing only interactions with compact support in 
 Minkowski space characterized by the choice of a test 
 function $g$. The ultraviolet problems show up in the 
 difficulty of defining the terms in the perturbative expansion 
(\ref{intfield:class}) of the 
 interacting quantum fields (the retarded products $R$),
 \begin{equation}
        A_{g{\mathcal{L}}_{1}}(f) = 
        R\bigl(\exp_{\otimes}(g{\mathcal{L}}_{1}),fA\bigr)
        \overset{\mathrm{def}}{=} 
        \sum_n \frac{1}{n!}R\bigl((g{\mathcal{L}}_{1})^
        {\otimes n},fA\bigr)\label{retprod} 
 \end{equation}
(where $A$ is an arbitrary polynomial in $\varphi$ and its derivatives)
 as everywhere defined distributions with values in ${\mathcal{W}}$.
On noncoinciding points they are already defined as symmetrized
iterated retarded Poisson brackets (\ref{Pb}).

 In the Bogoliubov-Epstein-Glaser approach 
 \cite{B:bos,B:eg,B:stora,B:scharf,B:bf} one expresses 
 them in terms of time ordered products. The time ordered 
 products can be inductively defined (this procedure is 
 equivalent to ultraviolet renormalization in approaches 
 via regularization) where the ambiguities are governed by 
 the renormalization group. It is somewhat cumbersome to 
 keep track of the $\hbar$-dependence during this 
 procedure (see \cite{B:df2}). Fortunately, there is an 
 alternative procedure mainly due to Steinmann\footnote{This book
relies on the LSZ-formalism, the retarded functions are the coefficients 
in Haag's series (which is an expansion of the interacting field in terms
of the free incoming fields). However, with some modifications the 
procedure works also in causal perturbation theory \cite{B:d}, which, in
contrast to the LSZ-framework, allows also for massless fields.}
\cite{B:steinmann} which works directly with the 
 retarded products. Namely, by (\ref{intfield:class}) they are causal,
\begin{equation}
 R\bigl((g{\mathcal{L}}_{1})^{\otimes n},f\mathcal{L}_{2}\bigr)=0 
 \quad\text{ for }\quad \mathrm{supp}\>g
\cap (\mathrm{supp}\>f+\bar V_-)=\emptyset\>,\label{causality}
\end{equation} 
symmetrical in the first entry and satisfy the relation
 \begin{gather}
         R\bigl(\exp_{\otimes}(g{\mathcal{L}}_{1}) 
         \otimes h{\mathcal{L}}_{2},f{\mathcal{L}}_{3}\bigr)
         - R\bigl(\exp_{\otimes}(g{\mathcal{L}}_{1}) 
         \otimes f{\mathcal{L}}_{3}, 
         h{\mathcal{L}}_{2} \bigr)\notag\\ 
         =\{ R\bigl(\exp_{\otimes}(g{\mathcal{L}}_{1}), 
         h{\mathcal{L}}_{2} \bigr), 
         R\bigl(\exp_{\otimes}(g{\mathcal{L}}_{1}),f{\mathcal{L}}_{3} 
         \bigr)\} ,\label{splitting}
 \end{gather}
which goes back to Lehmann-Symanzik-Zimmermann \cite{B:lsz}
and Glaser-Lehmann-Zimmermann \cite{B:glz}. (For a derivation
using Bogoliubov's definition of the interacting fields
 in terms of time ordered products, see \cite{B:df1}.)
 This relation holds as well in the classical as also in the 
 quantum case (note that suitable factors of $\hbar$ have 
 been included in the definition of the retarded products). 
 It also keeps the same form after performing the quotient 
 with the ideals ${\mathfrak{I}}^n$, hence it holds at any order in 
 $\hbar$. 
 
 From this relation we see that, as distributions in 
 $h\otimes f$, the terms of the left hand side are a 
 splitting of the right hand side into a retarded and an 
 advanced part. Therefore, the $n$-th order term is, in the 
 complement of the origin, uniquely determined by lower order 
 terms, and the extension to everywhere defined 
 distributions follows essentially the same path as in the 
Bogoliubov-Epstein-Glaser scheme \cite{B:steinmann,B:d}.
 
 In a last step we may now remove the restriction to 
 localized interactions, thereby solving the infrared 
 problem on a purely algebraic level. As above let $\mathcal{O}$
be a causally complete region. We consider a change of the interaction
outside of $\mathcal{O}$: $g\mathcal{L}_1\rightarrow (g\mathcal{L}_1+f
\mathcal{L}_2)$, $\mathrm{supp}\>f\cap\mathcal{O}=\emptyset$. We decompose
$f=f_+ +f_-$ with $\mathrm{supp}\>f_\pm\cap (\mathcal{O}+\bar V_{\mp})
=\emptyset$. Now let $h\in \mathcal{D}(\mathcal{O})$.
From (\ref{causality}) we know $A_{g\mathcal{L}_1+f_+
\mathcal{L}_2}(h)=A_{g\mathcal{L}_1}(h)$, and by using 
(\ref{retprod}), (\ref{causality}) and (\ref{splitting}) we obtain
the differential equation 
\begin{gather}
\frac{d}{d\epsilon}A_{g\mathcal{L}_1+\epsilon f_-\mathcal{L}_2}(h)
=\sum_n \frac{1}{n!}R\bigl((g\mathcal{L}_{1}+\epsilon f_-\mathcal{L}_2)^
        {\otimes n}\otimes f_-\mathcal{L}_2,hA\bigr)\notag\\
=\{\mathcal{L}_\epsilon ,
A_{g\mathcal{L}_1+\epsilon f_-\mathcal{L}_2}(h)\}\notag \>,
\end{gather}
where
\begin{displaymath}
        \mathcal{L}_\epsilon =\mathcal{L}_{2(g\mathcal{L}_{1}+
        \epsilon f_-\mathcal{L}_2)}(f_-).
 \end{displaymath}
It is solved by the Dyson series,
\begin{gather}
A_{g\mathcal{L}_1+f\mathcal{L}_2}(h)=
\sum_{r=0}^\infty\int_0^1 ds_r\int_0^{s_r}ds_{r-1}...\int_0^{s_2}ds_1
\{\mathcal{L}_{s_r} ,\{\mathcal{L}_{s_{r-1}},...
\{\mathcal{L}_{s_1} ,A_{g\mathcal{L}_1}(h)\}...\}\},\label{dyson}
\end{gather}
which is convergent in the sense of formal power series in the couplings.

We denote by $\mathfrak{A}_{g\mathcal{L}_1}(\mathcal{O})$ the algebra which 
is generated by the interacting fields $A_{g\mathcal{L}_1}(h),\>
h\in \mathcal{D}(\mathcal{O})$ and $A$ an arbitrary polynomial in
$\varphi$ and its derivatives. By means of (\ref{dyson}) one verifies that
$A_{g\mathcal{L}_1}(h)\rightarrow A_{g\mathcal{L}_1+
f\mathcal{L}_2}(h)$ is a canonical (classical field theory)
or unitary (quantum field theory) transformation which is independent 
of $A$ and $h$. Hence, $\mathfrak{A}_{\mathcal{L}_1}(\mathcal{O})$ may be,
as an abstract algebra, identified with $\mathfrak{A}_{g\mathcal{L}_1}
(\mathcal{O})$, where $g\in\mathcal{D}$, with $g\equiv 1$ on 
$\mathcal{O}$, is arbitrary. Embeddings $\mathfrak{A}_{\mathcal{L}_1}
(\mathcal{O}_1)\hookrightarrow \mathfrak{A}_{\mathcal{L}_1}(\mathcal{O})$
for $\mathcal{O}_1\subset\mathcal{O}$ are inherited from the inclusion
$\mathfrak{A}_{g\mathcal{L}_1}(\mathcal{O}_1)\subset 
\mathfrak{A}_{g\mathcal{L}_1}(\mathcal{O})$ (see \cite{B:df2})
and define a net of algebras which, according to the principles of
algebraic quantum field theory, characterizes the theory completely.
(For an alternative proof, which relies on the causal factorization of 
the time ordered products, see \cite{B:bfroma,B:bf} for the
quantum case and \cite{B:df2} for classical field theory.) 

Since the Poisson bracket on $\mathcal{V}$ (\ref{Pb}) is a power series
in $\hbar$, the canonical (unitary resp.) transformation (\ref{dyson})
respects the ideal $\mathfrak{I}$ of elements of order $\hbar$, thus
also the nets of algebras up to $n$ loops (\ref{A_n}) are well defined
(cf. \cite{B:df2}).
 \section{Conclusions and outlook}
We have shown that the conceptual frame of algebraic quantum field 
theory admits a clear formulation of the perturbative construction of
QFT. In particular the relation to classical field theory and to 
deformation quantization, as well as the role of the interpolating 
theories up to $n$ loops were clarified. One might hope that the latter 
theories even admit a nonperturbative construction.\footnote{We are 
grateful to G.Gallavotti for suggestions in this direction.}


\begin{thebibliography}{999}

\bibitem{B:bffls} Bayen, F.,Flato, M.,Fronsdal, C.,Lichnerowicz, A.,
Sternheimer, D. \textit{Deformation Theory and Quantization}, 
Annals of Physics (N.Y.) \textbf{111} (1978), p.61, p.111 
 
\bibitem{B:borchers1} Borchers, H. J. \textit{\"Uber die Mannigfaltigkeit
der interpolierenden Felder zu einer kausalen $S$-Matrix}, 
Nuovo Cimento  \textbf{15}  (1960), p.784.

\bibitem{B:borchers2} Borchers, H. J. \textit{Local rings and the connection 
of spin with statistics},   Commun. Math. Phys.  \textbf{1}  (1965), p.291.

\bibitem{B:bw} Bordemann, M. and Waldmann, S. \textit{Formal GNS 
construction and states in deformation quantization}, 
Commun. Math. Phys.  \textbf{195}  (1998), p.549.

\bibitem{B:bos} 
Bogoliubov, N. N., and Shirkov, D. V.: \textit{ Introduction to the 
Theory of Quantized Fiels}, John Wiley and Sons, 1976, 3rd edition.

\bibitem{B:bfroma} Brunetti, R., and Fredenhagen, K. 
\textit{Microlocal analysis and interacting quantum field theory: 
Renormalizability of $\varphi^4$}, in
\textit{Operator Algebras and Quantum Field Theory}, International Press 
(1997), Ed. Doplicher, S., Longo, R., Roberts, J. E., and Zsido, L.

\bibitem{B:bf} Brunetti, R., and Fredenhagen, K. 
\textit{Microlocal analysis and interacting quantum field theories: 
Renormalization on physical backgrounds}, Commun. Math. Phys. 
\textbf{208} (2000), p.623.  

\bibitem{B:bu} Buchholz, D. \textit{Algebraic Quantum Field Theory: 
A Status Report}, talk given at the XIIIth International Congress on 
Mathematical Physics, London (2000), math-ph/0011044

\bibitem{B:bufred} Buchholz, D. and Fredenhagen, K. \textit{Locality and
the structure of particle states}, Commun. Math. Phys., 
\textbf{84} (1982), p.1.  

\bibitem{B:bv} Buchholz, D. and Verch, R. \textit{Scaling algebras and 
renormalization group in algebraic quantum field theory}, Rev. Math.
Phys. \textbf{7} (1995), 1195

\bibitem{B:di} Dito, J. \textit{Star-Product Approach to Quantum Field Theory:
The Free Scalar Field}, Lett. Math. Phys. {\bf 20} (1990), p.125 

Dito, J., \textit{Star-products and nonstandard quantization for
Klein-Gordon equation}, J. Math. Phys. {\bf 33} (1992), p.791 

\bibitem{B:dhr} Doplicher, S., Haag, R. and Roberts, J. E. \textit{Fields, 
observables and gauge transformations I},  Commun. Math. Phys. 
\textbf{13} (1969), p.1.

Doplicher, S., Haag, R. and Roberts, J. E. \textit{Fields, observables
and gauge transformations II},  Commun. Math. Phys.
\textbf{ 15} (1969), p.173.

Doplicher, S., Haag, R. and Roberts, J. E. \textit{Local observables
and particle statistics I},  Commun. Math. Phys. 
\textbf{ 23} (1971), p.199.

Doplicher, S., Haag, R. and Roberts, J. E. \textit{Local observables
and particle statistics II},  Commun. Math. Phys. 
\textbf{ 35} (1974), p.49.

\bibitem{B:dr1} Doplicher, S. and Roberts, J. E. \textit{
Monoidal $C^*$-categories
and a new duality theory for compact groups},  Invent. Math. 
\textbf{ 98} (1989), p.157.

\bibitem{B:dr2} Doplicher, S. and Roberts, J. E. \textit{Why there is a 
field algebra with a compact gauge group describing the superselection
structure in particle physics}, Commun. Math. Phys. 
\textbf{ 131} (1990), p.51.

\bibitem{B:df1} D\"utsch, M., and Fredenhagen, K. \textit{A local 
(perturbative)
construction of observables in gauge theories: the example of QED},
Commun. Math. Phys. \textbf{ 203} (1999), p.71

\bibitem{B:df2} D\"utsch, M., and Fredenhagen, K. 
\textit{Algebraic quantum field theory, perturbation theory, and 
the loop expansion}, hep-th/0001129 (to appear in Commun. 
Math. Phys.)

\bibitem{B:df3} D\"utsch, M., Fredenhagen, K. and Boas, F.-M. 
\textit{The Master Ward Identity: Classical Field Theory}, work in preparation

\bibitem{B:d} D\"utsch, M. \textit{Causal perturbation theory in terms of 
retarded products}, work in progress

\bibitem{B:eg} Epstein, H., and Glaser, V. \textit{The role of locality in 
perturbation theory}, Ann. Inst. Henri Poincar\'e-Section \textbf{A XIX} 
n.3 (1973), p.211.

\bibitem{B:frs} Fredenhagen, K., Rehren, K.-H. and Schroer, B.
\textit{Superselection sectors with braid group statistics and exchange 
algebras I},  Commun. Math. Phys. \textbf{ 125} (1989), p.201.

\bibitem{B:glz} Glaser, V., Lehmann, H. and Zimmermann, W. \textit{Field
Operators and Retarded Functions}, Nuovo Cimen. \textbf{ 6} (1957), p.1122.

\bibitem{B:haag} Haag, R. \textit{ Local Quantum Physics: Fields, particles 
and algebras}, Springer-Verlag, Berlin, 2nd ed., 1996.

\bibitem{B:hk} Haag, R, and Kastler, D. \textit{An algebraic 
approach to field theory}, Journ. Math. Phys. \textbf{ 5}
(1964), p.848

\bibitem{B:lsz} Lehmann, H., Symanzik, K. and Zimmermann, W. \textit{On
the formulation of quantized field theories II},  Nuovo Cimen. 
\textbf{6} (1957), p.320.

\bibitem{B:micro} H\"ormander, L. \textit{ The Analysis of Linear Partial 
Differential Operators}, vol. I-IV, Springer-Verlag, 1983-1986.

\bibitem{B:peierls} Peierls,  R. \textit{The commutation laws of relativistic 
field theory}, Proc. Roy. Soc. (London) \textbf{ A 214} (1952), p.143

\bibitem{B:radman} Radzikowski, M. \textit{Micro-local approach 
to the Hadamard 
condition in quantum field theory on curved space-time}, Commun. 
Math. Phys., \textbf{ 179}, (1996), p.529

\bibitem{B:rehren} Rehren, K.-H. \textit{Local quantum 
observables in the AdS-CFT correspondence}, Phys. Lett. 
\textbf{ B493} (2000), p.383

\bibitem{B:scharf} Scharf, G. \textit{Finite Quantum Electrodynamics: 
The Causal Approach}, Springer-Verlag, 1995, 2nd edition.

\bibitem{B:steinmann} Steinmann, O. \textit{ Perturbation Expansions in 
Axiomatic Field Theory}, Lectures Notes in Physics 
\textbf{ 11}, Springer-Verlag, 1971.

\bibitem{B:stora} Stora, R. \textit{Differential algebras in 
Lagrangean field theory},
ETH Lectures, January-February 1993; and also, Popineau, S., and Stora, R. 
\textit{A pedagogical remark on the main theorem of 
perturbative renormalization theory}, unpublished preprint.



 
\end{thebibliography}
\end{document}